\documentstyle[12pt]{article}
\begin{document}
\thispagestyle{empty}
\begin{raggedleft}
UR-1401\\
ER-40425-261\\
hep-th/9511183\\
\end{raggedleft}
$\phantom{x}$\vskip 0.618cm\par
{\huge \begin{center}
A REMARK ON THE GAUGING OF CHIRAL BOSONS\footnote{This work is 
supported by CNPq, Bras\'\i lia, Brasil}
\end{center}}\par
\vfill
\begin{center}
$\phantom{X}$\\
{\Large Clovis Wotzasek\footnote{Permanent address:
Instituto de F\'\i sica, 
Universidade Federal do Rio de Janeiro,
Brasil}} \\[3ex] {\em
Department of Physics and Astronomy\\ University of Rochester\\ Rochester,
NY 14627 \\ USA}
\end{center}\par
\vfill
\begin{abstract}

\noindent We study the interacting chiral bosons and observe that a naive
gauging procedure
leaves the gauge invariant chiral constraint incompatible with the
field equations.  Consistency, therefore, rules out most gauging schemes:
in a left chiral scalar, only the left chiral current leads to a
consistent result,
in discordance with the present literature.  We use this gauging scheme
to show how the introduction of a chirally coupled gauge field 
becomes necessary to mod out the degree of freedom that obstructs
gauge invariance in a system of two chiral bosons, in order to sold
together right and left chiral bosons.
          
\end{abstract}
\vfill
\newpage

There has been a great amount of investigation on the proper way to 
gauge 2D self-dual fields\cite{siegel}\cite{florjack} in the last
few years
\cite{fs}\cite{sonn}\cite{bgp}\cite{lr}\cite{harada}\cite{bazeia}
\cite{gm1}\cite{gm2}\cite{clovis}\cite{nelson}, with discordant results. 
In the Floreanini-Jackiw model the basic difficulty seems to
come from the fact that this action is not manifestly Lorentz
covariant.
This makes the usual covariant derivative substitution meaningless. 
On the other hand, a direct coupling of the gauge
field with the matter current, say the vector one, which would
preserve gauge symmetry, results breaking Lorentz invariance\cite{gm1}. 
Despite these problems, the correct gauging of chiral bosons
has been achieved, although through some indirect ways. 
Bellucci et al\cite{bgp} and Harada\cite{harada} proposed to
project out
the free chiral component of the bosonized chiral Schwinger model,
leaving behind the other chiral component of the scalar field coupled
to the gauge potential.  Bazeia \cite{bazeia}
showed that
the same result follows from the linear constraint chiral
boson, which happens to be explicitly covariant.  Another indirect
proposal for the chiral boson gauging has been given by this author
\cite{clovis}
using a representation of the chiral field in terms of an infinite
set of self-coupled scalar fields\cite{wotzasek}.  In this covariant
representation one is allowed to use the covariant derivative
substitution.  Upon returning the gauged version to its chiral
representation, by elimination of the infinite set of auxiliary
Wess-Zumino fields, it results in an unambiguous coupling of chiral
bosons
to gauge fields.  This procedure has been extended
to the coupling of chiral bosons to gravitational fields\cite{nelson}.

Despite of the successes of these indirect gauging schemes for the
Floreanini-Jackiw model, it has been
reported that the explicitly covariant model for chiral bosons put forward
by Siegel\cite{siegel} also suffer from the same difficulties regarding
the coupling
to gauge fields.  This clearly shows that the trouble in doing this
coupling is
less related to the Lorentz covariance, as reported, than to the chiral
nature of the
constraints in these models.  Unfortunately, we do not have the
indirect gauging schemes for Siegel model as we do for the FJ model:
while the FJ model can be obtained from the action
of a free scalar field by application of a chiral
projector, the model proposed by Siegel is 
obtained from the action of a scalar field on a gravitational background
by truncation of the original metric.  The
truncation process preserves the nature of the original coupling,
while the chiral projector
changes a vector coupling into a chiral coupling.

In this Letter we show that, in a (left) chiral boson theory,
the coupling of the gauge potential with axial, vector or right chiral
currents leaves the gauge invariant constraint incompatible
with the field equations, being obstructed by the
presence of the gauge anomaly.  The appearance of an anomaly is expected
since bosonic models can be thought as effective theories, at
one-loop level, for the corresponding fermionic models.  We propose to
use compatibility as a guiding rule to choose the proper
gauging procedure among the different possibilities.  Based on this,
we then show that all
covariant derivative substitution schemes fail to satisfy this consistency
check, both for the FJ and Siegel's models.  It is worth noting that in order
to correctly implement these couplings
we have to look at the chiral theories as models for scalar particles
in a gravitational background.  This seems necessary to
correctly define the vector current as dual to the axial current since
for abelian scalar fields one does not have a direct way to compute the
vector current as a Noether current.
This point seems to have been overlooked in some of the past studies
of the subject or, the definition of the dual Hodge transformation, used to
construct the vector current from the
axial one, was done incorrectly.  We end this Letter with a simple application
showing how the introduction of a gauge field, chirally coupled to the chiral
matter, becomes necessary in order to sold
together a right and a left chiral boson, using a technique presented
a few years ago by Stone\cite{stone}.

The basic difficulty in the gauging of chiral boson has been pointed
out recently by Balachandran et al\cite{bala}, in the context of quantum
Hall effect.   Take, for instance, the
action for a flat space-time free scalar field in the light-front
coordinates,

\begin{equation}
{\cal L}= \partial_+\phi\partial_-\phi
\end{equation}

\noindent The chiral constraint $\partial_-\phi\approx 0$ is consistent
with the equations of motion before gauging, but not after.  Indeed,
suppose we gauge the system through the chiral derivative substitution
rule\footnote{Our notation is as follows: $x^{\pm}=
{1\over{\sqrt 2}}(x^0\pm x^1)$ ; $\epsilon^{+-}=\epsilon^{-+}=1$.}

\begin{eqnarray}
\partial_+\phi & \longrightarrow & \partial_+\phi\nonumber\\
\partial_-\phi & \longrightarrow & D_-\phi
\end{eqnarray}

\noindent where $D_-\phi=\partial_-\phi + A_-$.  The equation of motion
for the scalar field now reads

\begin{equation}
\partial_+D_-\phi \sim E
\end{equation}

\noindent with $E$ being the electric field on the line.  Notice that
the imposition of the gauge invariant constraint ($D_-\phi\approx 0$)
becomes inconsistent with the field equation due to the appereance
of the gauge anomaly.  This picture is in fact quite general: suppose that
the equations of motion, before gauging, are giving by

\begin{equation}
\label{compat1}
L\:\partial_-\phi=0
\end{equation}

\noindent  with $L$ being the appropriated Euler-Lagrange differential
operator.  If the gauging procedure chosen is the simple substitution
of partial
derivatives by their covariant counterparts, the outcome
after gauging will be

\begin{equation}
\label{compat2}
L\:D_-\phi\sim E
\end{equation}

\noindent  showing the above mentioned inconsistency.  One may ask if, for
chiral bosons, there exists any gauging procedure in which the chiral
constraint remains compatible after turning on the interactions.  The positive
answer is given by the action proposed in Refs.\cite{bgp}
and \cite{harada} for the Floreanini-Jackiw model,obtained through the
application of the chiral projector on the chiral Schwinger model. In
fact, from the Lagrangian density proposed by these authors

\begin{equation}
{\cal L}= \partial_+\phi\partial_-\phi-\partial_-\phi\partial_-\phi +
2e(\partial_+\phi -\partial_-\phi) A_- +\mbox{Gauge Terms}
\end{equation}

\noindent we get $\partial_1D_-\phi=0$ as its equation of motion,
which shows that the imposition of gauge invariant chiral
constraint is not obstructed by the gauge anomaly.  

What we need to do next is
to find out what is the direct gauging scheme leading to this action. 
To begin with, let us consider the case of the scalar
field coupled to background gravitational field.  FJ and Siegel models
for chiral bosons can be considered as belonging to this class of
theories with some
special choice of metric.  The action for the standard minimal coupling of
a scalar field to a metric $g_{\mu\nu}$ is

\begin{equation}
\label{geometric}
{\cal L}_0={1\over 2} \sqrt{-g} g^{\mu\nu}\partial_\mu\phi
\partial_\nu\phi
\end{equation}

\noindent A convenient parametrization of the metric is given as
(notice that it does not correspond to a partial gauge-fixing but
is consequence of the Weyl symmetry)

\begin{equation}
\label{metric}
\sqrt{-g} g^{\mu\nu}={1\over{1-\lambda_{++}\lambda_{--}}}\left(
\begin{array}{cc}
{2\lambda{--}} & {1+ \lambda_{++}\lambda_{--}}\\
{1+\lambda_{++}\lambda_{--}} & {2\lambda_{++}}
\end{array}
\right)
\end{equation}

\noindent The (left) FJ and Siegel models can be obtained simply by
truncation of this metric as ($\lambda_{--}=0$) for Siegel, and
($\lambda_{--}=0\:,\:\lambda_{++}=-1$) for FJ.  The axial current is

\begin{eqnarray}
J_{(A)}^+ &=& {1\over{1-\lambda_{++}\lambda_{--}}}\left[2\lambda_{--}
\partial_+\phi +(1+\lambda_{++}\lambda_{--})\partial_-\phi\right]
\nonumber\\
J_{(A)}^- &=& {1\over{1-\lambda_{++}\lambda_{--}}}\left[2\lambda_{++}
\partial_-\phi +(1+\lambda_{++}\lambda_{--})\partial_+\phi\right]
\end{eqnarray}

\noindent The vector current is defined as dual to the axial one
\footnote{Defining the vector current as dual to the axial current is not
really a restriction of our method since this is a feature of the
two dimensions.  In any case, one can show that in the non-abelian case,
where the vector current can be defined as a Noether current, everything
works as discussed here.}

\begin{equation}
J_{(V)}^\mu=\mbox{}^*J^\mu_{(A)}
\end{equation}

\noindent where the usual Hodge transformation must be generalized to

\begin{equation}
\mbox{}^*J^\mu_{(A)}=\sqrt{-g} g^{\mu\nu}
\epsilon_{\nu\lambda}J_{(A)}^\lambda
\end{equation}

\noindent  in order to take into account the presence of the gravitational
background.  A simple calculation shows that

\begin{eqnarray}
\label{vector}
J^+_{(V)} &=& -\:\partial_-\phi\nonumber\\
J^-_{(V)} &=& \:\partial_+\phi
\end{eqnarray}

\noindent that is topologically conserved, as it should.  Observe that
the vector current is metric independent, being the same for the chiral
models defined by truncation of the metric.

We are now in position to compute the right and left chiral currents.  We find,

\begin{eqnarray}
\label{left}
J^+_{(L)} &=& {2\lambda_{--}\over{1-\lambda_{++}\lambda_{--}}}
\left(\lambda_{++}\partial_-\phi+\partial_+\phi\right)\nonumber\\
J^-_{(L)} &=& {2\over{1-\lambda_{++}\lambda_{--}}}
\left(\lambda_{++}\partial_-\phi+\partial_+\phi\right)
\end{eqnarray}

\noindent for the left current, and

\begin{eqnarray}
J^-_{(R)} &=& {2\lambda_{++}\over{1-\lambda_{++}\lambda_{--}}}
\left(\lambda_{--}\partial_+\phi+\partial_-\phi\right)\nonumber\\
J^+_{(R)} &=& {2\over{1-\lambda_{++}\lambda_{--}}}
\left(\lambda_{--}\partial_+\phi+\partial_-\phi\right)
\end{eqnarray}

\noindent for the right current.  Observe that the (left) chiral boson
limit ($\lambda_{--}=0$) kills the $J^+_{(L)}$ component, leaving
the left current holomorphically conserved, while the right chiral
current has both components non-vanishing.  This is certainly a
desired result for the chiral case.  In fact, in the flat space-time
theory for the free scalar field, there are two separated affine
invariances for the left and right chiral sectors.  These symmetries
are reflected by the fact that both the right and left chiral currents have
only one non-zero component, $J_{(L)}^-=J_{(L)}^-(x^+)$ and
$J_{(R)}^+=J_{(R)}^+(x^-)$, since

\begin{eqnarray}
\partial_-J_{(L)}^- &=& 0\nonumber\\
\partial_+J_{(R)}^+ &=& 0
\end{eqnarray}

\noindent and generate two commuting affine algebras.  However, in the
chiral case, only one of these currents keeps this property, i.e., either

\begin{eqnarray}
J_{(L)}^- &=& J_{(L)}^-(x^+) \;\;\;\;\mbox{but}\;\;\;\;J_{(R)}^+\neq
J_{(R)}^+(x^-)\nonumber\\
&\mbox{or}&\nonumber\\
J_{(L)}^- &\neq& J_{(L)}^-(x^+) \;\;\;\;\mbox{but}\;\;\;\;J_{(R)}^+ =
J_{(R)}^+(x^-)
\end{eqnarray}

\noindent This can also be seen from the fact that, for chiral theories,
while the vector and axial transformations are global symmetries, the
affine transformations are semi-local symmetries.  Take for instance the
case of a left Siegel boson.  The semi-local shift $\phi\rightarrow
\phi +\alpha(x^+)$ certainly leaves the action invariant, but
$\phi\rightarrow \phi +\alpha(x^-)$ does not.  The Noether current is
immediately identified as

\begin{equation}
J^- =2(\partial_+\phi+\lambda_{++}\partial_-\phi)\;\;\;\mbox{,}\;\;\;
J^+=0
\end{equation}

\noindent which is easily seen to be the (left) chiral limit of (\ref{left}).

We shall examine next the coupling with an external electromagnetic field. 
We do this iteratively,
introducing the necessary Noether counter-terms.  First we examine the
coupling with the vector current (\ref{vector}): 
${\cal L}_0\rightarrow {\cal L}_0 + A_+J^+_{(V)}+A_-J^-_{(V)}$ where
${\cal L}_0$ is defined by Eq.(\ref{geometric}).  We observe
that, after gauging, the vector current remains conserved, but
the axial current does not.  In fact a direct calculation shows

\begin{equation}
\partial_\mu J^\mu_{(A)}=\partial_-A_+-\partial_+A_-=E
\end{equation}

\noindent  which, again, being independent of the metric elements,
is valid for all cases.  We observe that
before gauging the chiral constraint cannot be imposed 
compatibly, differently from what happens in the flat space-time
case.  However, if we restrict ourselves to the chiral models, where
$\lambda_{--}=0$, 
then we have compatibility for the free theory (see Eq.\ref{compat1}). 
This consistency is destroyed for the gauging with vector
currents (Eq.\ref{vector}) due to the anomaly.  Therefore, we have
to rule out the vector current gauging as inappropriate for chiral theories.

Let us consider next the case of axial coupling.  The free action changes
to 

\begin{equation}
{\cal L}_0\rightarrow{\cal L}_1={\cal L}_0 + A_+J^+_{(A)}+A_-J^-_{(A)}
\end{equation}

\noindent Under an axial transformation the action ${\cal L}_1$ does
not remain invariant but its variation is given by

\begin{equation}
\delta{\cal L}_1=-\delta\left\{{\lambda_{--}\over{1-\lambda_{++}\lambda_{--}}}
A_+^2 + {\lambda_{++}\over{1-\lambda_{++}\lambda_{--}}} A_-^2 + 
{{1+\lambda_{++}\lambda_{--}}\over{1-\lambda_{++}\lambda_{--}}}A_+A_-\right\}
\end{equation}

\noindent  Therefore, a further modification of the action as

\begin{equation}
{\cal L}_1\rightarrow{\cal L}_2={\cal L}_1 + {\lambda_{--}
\over{1-\lambda_{++}\lambda_{--}}}\left[\lambda_{--}A_+^2+\lambda_{++}A^2_-
+\left(1+\lambda_{++}\lambda_{--}\right)A_+A_-\right]
\end{equation}

\noindent leaves the action invariant under an axial transformation. 
It is simple to check that here, the axial current
remains conserved, while the gauge coupling modifies the vector current,
which now fails to be conserved.  Finally, by checking the equations of
motion, one notices the incompatibility of the gauged constraint with the
field equation, which also
rules out this coupling as inappropriate for the chiral models.

We are then left only with the possibilities of chiral current couplings. 
We have explicitly checked that for left chiral bosons, the coupling with the
right chiral current results incompatible.  Let us work out explicitly
the coupling of the left chiral current and verify
that this is the only possible consistent way of coupling.  The Noether
procedure then gives

\begin{equation}
{\cal L}_0\rightarrow{\cal L}_1={\cal L}_0 + A_+J^+_{(L)}+A_-J^-_{(L)}
\end{equation}

\noindent whose variation reads

\begin{equation}
\delta{\cal L}_1 = {1\over{1-\lambda_{++}\lambda_{--}}}\left\{
\left(2\lambda_{++}\lambda_{--}A_+\partial_-\alpha + 2A_-
\partial_+\alpha\right)-\delta\left(\lambda_{--}A_+^2+
\lambda_{++}A_-^2\right)\right\}
\end{equation}

\noindent The second term can be reabsorbed into a redefinition of
the action as

\begin{equation}
\label{axvar}
{\cal L}_1\rightarrow{\cal L}_2={\cal L}_1 +\lambda_{++}A_-^2 +
\lambda_{--}A_+^2
\end{equation}

\noindent but the first piece cannot be eliminated by any choice of
a Noether counter-term.  This is true even for the truncated chiral
limit.  However, this action
has the nice property of having its variance independent of the 
matter fields.  This property will be explored next in order to sold
together two bosons of opposite chiralities. 
The reader can also observe that the truncation process
used to go from the non-chiral to the chiral case does not change the
nature of the coupling, which means that the vector, axial and chiral
couplings studied above remains the same after truncation.  This is
certainly different from the projection process using chiral constraints
that transform, for instance, the vector coupling into a chiral
coupling.

Let us examine more closely the compatibility of the gauged chiral
constraint with the equations of motion.  Before gauging, the equation
of motion for the matter fields read

\begin{equation}
0 = \partial_+\left({2\lambda_{--}\over{1-\lambda^2}}
\partial_+\phi + {{1+\lambda^2}
\over{1-\lambda^2}}\partial_-\phi\right)
+ \partial_-\left({2\lambda_{++}\over{1-\lambda^2}}
\partial_-\phi + {{1+\lambda^2}
\over{1-\lambda^2}}\partial_+\phi\right)
\end{equation}

\noindent where $\lambda^2=\lambda_{++}\lambda_{--}$. 
Notice that the chiral
constraint is not compatible with the equations of motion
before gauging, however, restriction to the chiral
limit gives

\begin{eqnarray}
0 &=& (\partial_+ + \partial_-\lambda_{++} 
+\lambda_{++}\partial_-)\partial_-\phi \;\;\;\;\mbox{(Siegel)}\nonumber\\
0 &=& (\partial_+ - \partial_-)\partial_-\phi \;\;\;\;\mbox{(FJ)}
\end{eqnarray}

\noindent which are compatible with $\partial_-\phi\approx 0$.  After gauging
with (left) chiral currents, the result is

\begin{eqnarray}
0 &=& (\partial_+ + \partial_-\lambda_{++} 
+\lambda_{++}\partial_-)D_-\phi \;\;\;\;\mbox{(Siegel)}\nonumber\\
0 &=& (\partial_+ - \partial_-)D_-\phi \;\;\;\;\mbox{(FJ)}
\end{eqnarray}

\noindent We see, as discussed in the introduction, that for (left)
chiral couplings, the
gauge invariant constraint can be imposed over the field equations
without being obstructed by the gauge anomaly, and that this is
the only consistent possibility for the (left) chiral boson.

As a simple application, let us describe how to sold together two
Floreanini-Jackiw chiral bosons of opposite
chiralities\cite{stone}\cite{amorim}.  This is non trivial since
the sum of the actions of a right and a left chiral boson is not equal
to the action of a single scalar field.  This should be clear since
these two objects possess two opposite individual affine symmetries
that cannot
be combined into a vector symmetry since these fields do not belong
to the same multiplet.  To remove the obstruction to gauge
symmetry one has to introduce a gauge field that absorbs the phase
difference for the two independent transformations.  Besides, if we
do not introduce a kinetic term for the gauge field, it can be
integrated
out in the path integral, leaving behind its effect over each chiral
component.  This selects for survival only those
configurations where the right and the left fields belong to the same
multiplet, effectively soldering them together.

For computational convenience we use front-form variables. 
In this coordinate system, the action for left and right FJ chiral
bosons read, respectively

\begin{equation}
{\cal L}_0^{(\pm)} = \mp\dot\phi_{\pm}\phi_{\pm}^\prime -
(\phi_{\pm}^\prime)^2
\end{equation}

\noindent where dot and prime have their usual significance as time and
space derivatives.  We know, from their field equations, that these
models have a residual
invariance under a semi-local transformation

\begin{equation}
\phi_{\pm}\rightarrow\tilde\phi_{\pm}=\phi_{\pm} +\alpha_{\pm}(t)
\end{equation}

\noindent  Therefore, if one defines a scalar field as a combination
of these chiral ones as

\begin{equation}
\label{scalar}
\Phi = \phi_- - \phi_+
\end{equation}

\noindent then clearly the combination of the two semi-local
transformations above will not lead to a vector transformation for the
scalar field, unless some constraint is imposed over each individual
component.  This is the role played by the gauge field.  We can then
follow the gauging procedure described above to obtain the
action of an interacting chiral boson coupled to a gauge field through
their chiral left and right currents, respectively

\begin{equation}
{\cal L}_0^{\pm}\rightarrow {\cal L}_1^{\pm}={\cal L}_0^{\pm}
\:\mp 2A\left(\dot\phi_{\pm}\pm \phi_{\pm}^\prime\right)
\end{equation}

\noindent Here $A$ is the space-component of the gauge field.  Although
each individual (gauged) action is variant under a gauge
transformation, as we have seen in the last section, (see Eq.\ref{axvar})
one can verify that

\begin{equation}
{\cal L}={\cal L}_1^- + {\cal L}_1^+ - 2 A^2
\end{equation}

\noindent is indeed invariant.  The last term is a contact term that
compensates for
the non-invariances of each chirality.  Now, we can integrate out the
gauge field $A$, as

\begin{equation}
e^{i\int d^2 x {\cal W}}=\int[dA]e^{i\int d^2 x {\cal L}}
\end{equation}

\noindent to obtain

\begin{equation}
{\cal W}={1\over 2}\partial_\mu\Phi\partial^\mu\Phi
\end{equation}

\noindent which is the action for the scalar field (\ref{scalar})
defined as a combination of a right and a left chiral boson.

In conclusion, in this work we have studied the problem of coupling self-dual scalar
fields in 2D to an external electromagnetic field described by a vector
potential $A_\mu$.  We have recognized the basic difficulty
as an incompatibility between the gauge invariant
chiral constraint and the field equation for the matter field.  Using
consistency as a guiding rule, we have worked out the
coupling of gauge fields with different matter currents and, observed
that the only consistent coupling for a left chiral matter is
with a left chiral current.  This explain the results obtained  with the
use of the chiral projector on the Schwinger model.  As an application
we verified that in order to sold together a left and a right chiral
boson into a scalar field, we must (chirally) couple them to a gauge field that
will mod out the degree of freedom that obstructs gauge
invariance.\vspace{0.1cm}\\

\noindent ACKNOWLEDGMENTS:  We want to thank Drs. R.Amorim and A.Das for
many useful suggestions and comments.\vspace{0.1cm}\\

\end{document}